\documentclass[sigconf,nonacm]{acmart}
\usepackage{tikz}
\usetikzlibrary{arrows.meta,positioning,fit,calc}
\usepackage{pgfplots}
\pgfplotsset{compat=1.18}
\usepackage{float}

\setcopyright{none}
\acmConference[KDD Cup 2026 UniRec Workshop]
{KDD Cup 2026 Tencent UniRec Challenge Workshop}
{August 12, 2026}
{Jeju, Korea}
\acmYear{2026}
\acmDOI{}
\acmISBN{}

\AtBeginDocument{%
  }

\newcommand\blfootnote[1]{%
  \begingroup
  \renewcommand\thefootnote{}%
  \footnotetext{#1}%
  \addtocounter{footnote}{-1}%
  \endgroup
}

\title{UniDot: A Unified Network for Sequence Modeling and Feature Interaction in Large-scale Recommendation}

\author{Rongcheng Lin}
\affiliation{%
  \institution{Meta}
  \city{Menlo~Park}
  \country{USA}
}
\email{linrongc@meta.com}

\author{Yan Sun}
\affiliation{%
  \institution{Meta}
  \city{Menlo~Park}
  \country{USA}
}
\email{yansun@meta.com}

\author{Jamey Zhang}
\affiliation{%
  \institution{Meta}
  \city{Menlo~Park}
  \country{USA}
}
\email{jameyz@meta.com}

\author{Guanglei Xiong}
\affiliation{%
  \institution{Meta}
  \city{Menlo~Park}
  \country{USA}
}
\email{glx@meta.com}

\author{Ivan Ji}
\affiliation{%
  \institution{Meta}
  \city{Menlo~Park}
  \country{USA}
}
\email{ivanji@meta.com}

\author{Xianjie Chen}
\affiliation{%
  \institution{Meta}
  \city{Menlo~Park}
  \country{USA}
}
\email{cxj@meta.com}

\author{Shujian Bu}
\affiliation{%
  \institution{Meta}
  \city{Menlo~Park}
  \country{USA}
}
\email{shujian@meta.com}

\renewcommand{\shortauthors}{Lin et al.}

\begin{document}

\begin{abstract}
Industrial recommenders rely on two model families that have evolved largely
independently: feature-interaction models over multi-field user/item features,
and sequential models over user-behavior histories. Production systems couple
them only loosely. To unify the two, we present
\textbf{UniDot}, a novel architecture for
post-click conversion prediction built from the \emph{factorization-machine (FM)
point of view}: the embedding inner product---which powers collaborative
filtering and lets a recommender generalize to unseen user--item pairs---is the
same primitive as attention's query$\cdot$key scoring, so a single
\textbf{dot-product of tokens} can underlie both feature interaction and sequence
modeling. UniDot tokenizes non-sequential fields and multi-domain behavioral
sequences into one shared token space and stacks a single macro-block in which a
\textbf{token-mixing bus} and a \textbf{sequence-retrieval bus} (item tokens
cross-attending the histories) run \emph{in parallel} and exchange state each
layer through an MLP-Mixer fusion, while an \textbf{FM Highway} carries explicit
per-layer dot-product interactions \emph{around} the residual stack directly to
the classifier. The sequence side is embedded once per forward pass and shared by
all consumers, bounding inference latency. Trained with a dual sparse/dense
(Adagrad + Muon) optimizer, an auxiliary conversion-delay head, and multi-path
mutual learning, UniDot finished as the \textbf{runner-up} on the
\textbf{Industrial track} of the TAAC $\times$ KDD Cup 2026.
\end{abstract}

\keywords{recommender systems, feature interaction, sequence modeling,\\ CTR/CVR prediction, token mixing, factorization machines}

\maketitle
\blfootnote{\scriptsize
KDD Cup 2026 Tencent UniRec Challenge Workshop, August 12, 2026, Jeju, Korea.\\
Competition website: \url{https://algo.qq.com/}.\\
All experiments, data collection, and processing activities were conducted on
the competition host's infrastructure. No experiments, data collection or
processing activities were conducted on Meta infrastructure.
}

\section{Introduction}
Large-scale recommendation systems rank enormous candidate sets in real time and
underpin modern content and advertising platforms. Their predictive models have
matured along two largely separate traditions.
\textbf{(1) Feature-interaction models}---Wide\&Deep \cite{cheng2016},
DeepFM \cite{guo2017}, xDeepFM \cite{lian2018},
DCN/DCN-v2 \cite{wang2017dcn,wang2021}, FiBiNet \cite{huang2019},
AutoInt \cite{song2019}, Wukong \cite{zhang2024wukong},
DHEN \cite{zhang2022dhen}---learn explicit and implicit crosses among a wide,
mostly \emph{static} feature set (user profile $\times$ item attributes).
\textbf{(2) Sequential user-interest models}, e.g.\ DIN \cite{zhou2018},
DIEN \cite{zhou2019dien}, DSIN \cite{feng2019dsin}, SIM \cite{pi2020},
ETA \cite{chen2021eta}, TWIN \cite{chang2023}, LONGER \cite{chai2025longer},
and TIN \cite{zhou2024tin}, model the \emph{dynamics} of user behavior, typically
with target-aware attention over a single behavior history.

The \textbf{TAAC $\times$ KDD Cup 2026 (Tencent Uni-Rec Challenge)}---\emph{``Towards
Unifying Sequence Modeling and Feature Interaction for Large-scale
Recommendation''}---targets exactly this gap, asking for a \textbf{unified
tokenization scheme} and a \textbf{homogeneous, stackable backbone} that models
sequential and non-sequential features in one architecture for large-scale
recommendation, ranked by AUC under an inference-latency budget (\S\ref{sec:data}
describes the data). Following the recently introduced SlimPer
framework~\cite{slimper2026}, we designed UniDot. It is a direct answer to this
brief: one stackable block, one shared token space, unifying both families. The
name abbreviates
\textbf{Uni}fied modeling via \textbf{D}ot-products \textbf{O}f
\textbf{T}okens---cross-attention (sequence retrieval) and factorization machines
(feature interaction) are both dot-products of tokens, the single primitive the
whole model is built on.

This vantage point is deliberate. The FM inner product
$\langle v_u, v_i\rangle$ is the engine of \emph{collaborative filtering}: user
and item meet through latent factors estimated from all observed
co-occurrences, so the score generalizes to user--item pairs never seen together
in training~\cite{rendle2010}---and in advertising conversion data, with sparse
positives over a large, fast-moving ad inventory, most candidate pairs at serving
time are effectively new. Prior unification work makes this inner product
\emph{implicit}---an emergent property of a deep interaction stack or
transformer. UniDot keeps it explicit wherever signal crosses the user--item or
candidate--history boundary, using the deeper machinery (token mixing, gated
MLPs, attention) to refine the \emph{operands} of those dot-products rather than
replace the operation.

\begin{figure*}[t]
\centering
\resizebox{\textwidth}{!}{%
\begin{tikzpicture}[
  font=\footnotesize,
  >={Stealth[length=2.2mm]},
  box/.style={draw, rounded corners=2pt, align=center, inner sep=3pt, minimum height=6mm},
  inp/.style={box, fill=gray!12},
  tokz/.style={box, fill=green!14},
  bus/.style={box, fill=orange!22, minimum width=56mm},
  fuse/.style={box, fill=yellow!30},
  cls/.style={box, fill=blue!14, minimum width=26mm},
  hwl/.style={->, red!65!black, thick},
]
\node[inp] (profile) at (0,1.15) {user / item\\ profile fids\\ + pre-trained embs};
\node[inp] (seqs) at (0,-1.15) {behavioral\\ domains $S_1..S_4$};
\node[tokz] (tok) at (3.0,1.15) {tokenizers (\S\ref{sec:tok})\\ $\mathbf{U},\mathbf{I},\mathbf{I}_h$};
\node[tokz] (pipe) at (3.0,-1.15) {seq pipeline (\S\ref{sec:seqenc})\\ \emph{embed once, share}\\ $\mathbf{H}^{(1)}..\mathbf{H}^{(S)}$};
\node[bus] (mix) at (9.4,1.15) {\textbf{token-mixing bus}\\ Wukong / TokenMixer $\times W$};
\node[tokz] (pl) at (6.9,0.0) {MultiChannel\\ SeqPool};
\node[fuse] (fu) at (10.6,0.0) {\textbf{MLP-Mixer fusion}\\ (FuseFFN)};
\node[bus] (ret) at (9.4,-1.15) {\textbf{sequence-retrieval bus}\\ $\mathbf{I}_h$ cross-attends each $\mathbf{H}^{(s)}$ + fusion FFN};
\node[draw, dashed, rounded corners=4pt, inner sep=2.0mm,
      fit=(mix)(pl)(fu)(ret),
      label={[font=\footnotesize\itshape, anchor=south east]north east:macro-block $\times L$}] (blk) {};
\node[cls] (clf) at (15.4,0.0) {\textbf{classifier}\\ $\rho$ + MLP $\to \hat y$\\ {\scriptsize + aux delay head}};
\draw[->] (profile) -- (tok);
\draw[->] (seqs) -- (pipe);
\draw[->] (tok.east) -- node[above]{\scriptsize $[\mathbf{U};\mathbf{I}]$} (mix.west);
\draw[->] ([yshift=-3mm]tok.east) -- ++(0.5,0) |-
  node[left, pos=0.22]{\scriptsize $\mathbf{I}_h$} ([yshift=3mm]ret.west);
\draw[->] (pipe.east) -- (ret.west);
\draw[->] (pipe.east) -- node[above, pos=0.72, sloped]{\scriptsize views} (pl.west);
\draw[->] (pl.north) -- node[right, pos=0.4]{\scriptsize pooled} (pl.north |- mix.south);
\draw[<->] (fu.north) -- (fu.north |- mix.south);
\draw[<->] (fu.south) -- (fu.south |- ret.north);
\draw[->] (mix.east) -- node[above, pos=0.35]{\scriptsize $Z^{L}_{\text{mix}}$} (mix.east -| clf.west) -- (clf.160);
\draw[->] (ret.east) -- node[below, pos=0.35]{\scriptsize $Z^{L}_{\text{seq}}$} (ret.east -| clf.west) -- (clf.200);
\draw[hwl] (ret.south) -- ++(0,-0.55) -| node[above, pos=0.18]
  {\scriptsize \textbf{FM Highway}: $[\phi^{1};\dots;\phi^{L}]$ (dots + Gram + cross-dots), bypasses fusion}
  (clf.south);
\draw[->] (profile.north) -- ++(0,0.3) -| node[above, pos=0.25]
  {\scriptsize emb-skip signal $e_{\text{skip}}$} (clf.north);
\end{tikzpicture}%
}
\caption{UniDot architecture (\S\ref{sec:method}): two buses co-evolve through
$L$ macro-blocks, exchange state via the per-layer MLP-Mixer fusion, and the FM
Highway routes explicit per-layer dot-products past the fuser to the classifier.}
\label{fig:arch}
\end{figure*}
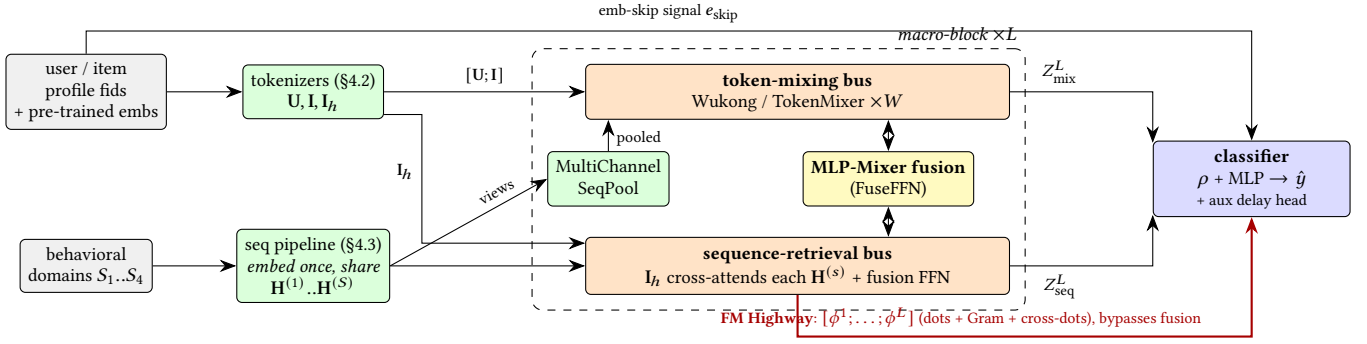

UniDot stacks one homogeneous block $L$ times: two buses---a \textbf{token-mixing
bus} over user/item profile tokens and a \textbf{sequence-retrieval bus} whose
item tokens cross-attend the behavioral histories---exchange state each layer
through an \textbf{MLP-Mixer fuser}, while an \textbf{FM Highway} routes explicit
dot-product interactions straight to the classifier. Everything else
(\S\ref{sec:method}) is \emph{how} these pieces are made cheap, candidate-aware,
and scalable. Around this design---plus the training recipe behind our final
entry---we contribute four ideas.
\textbf{(1) A dual-path, per-layer-parallel block:} rather than cascading a
sequence module into an interaction backbone, each block runs two paths
\emph{concurrently}---a non-sequential \textbf{token-mixing} path (a swappable
mixer slot: Wukong's parallel LCB + FMB by default, a TokenMixer-style block as a
drop-in) and a \textbf{sequence-retrieval} path (item tokens cross-attend the
behavioral sequences)---which exchange state every layer via a canonical
MLP-Mixer fusion (\S\ref{sec:perlayer}), so profile and sequence signal co-evolve;
this interleaving is \textbf{inspired by} InterFormer~\cite{zeng2024} and
Kunlun~\cite{kunlun}, from which UniDot differs in keeping the cross-boundary
signal an \emph{explicit dot-product}.
\textbf{(2) A shared, candidate-aware, multi-domain sequence pipeline:} all
behavioral domains are embedded once and reshaped to a \emph{uniform per-position
width} by a position-local fid-axis compression (\S\ref{sec:tok}); a DIN-style
\textbf{conditionally gated SwiGLU} injects candidate context into \emph{only the
gate}, and a timestamp-interleaved \textbf{merged stream} models cross-domain
temporal patterns (\S\ref{sec:seqenc}).
\textbf{(3) The FM Highway for explicit feature interaction:} per-domain
query--key dot products, an aggregated Gram matrix, and cross-bus user--item dots
are concatenated across layers and fed \emph{directly} to the classifier
(\S\ref{sec:perlayer}), bypassing the residual fusion path and preserving the
FM-style second-order signal a deep residual stack tends to wash out, at
negligible cost.
\textbf{(4) Multi-path mutual learning over shared embeddings:} we train
\textbf{two} UniDot paths jointly on one shared (and dominant) embedding table,
each regularized toward the other's predictions (\S\ref{sec:dml}), pulling even a
\emph{single} served path to a better minimum---a 1$\times$-cost model that would
itself place as runner-up (Table~\ref{tab:leaderboard})---while the two-path mean
recovers the small remainder.

UniDot
\textbf{finished as the runner-up on the Industrial track} of the TAAC $\times$ KDD Cup
2026 challenge under the official AUC metric,
reaching a final-leaderboard AUC of \textbf{0.83217} (Table~\ref{tab:leaderboard}). This result came from
\emph{architecture and scale} rather than heavy feature engineering: we add no
hand-crafted cross features beyond the released schema.

\begin{table}[t]
\caption{Final Industrial-track leaderboard (round 2, top 10). $^\star$Serving a
\emph{single} path of our all-data DML model (1$\times$ inference cost) would
still rank second---multi-path training pulls even a single model to a better
minimum (\S\ref{sec:dml}).}
\label{tab:leaderboard}
\centering
\small
\begin{tabular}{clc}
\toprule
Rank & AUC $\uparrow$ & Gap to \#1 \\
\midrule
1 & 0.83254 & --- \\
\textbf{2 (UniDot, ours)} & \textbf{0.83217} & $0.037\%$ \\
\textit{\, (UniDot, single path)}$^\star$ & \textit{0.83184} & $0.070\%$ \\
3 & 0.83145 & $0.109\%$ \\
4 & 0.83080 & $0.174\%$ \\
5 & 0.83073 & $0.181\%$ \\
6 & 0.83036 & $0.218\%$ \\
7 & 0.82915 & $0.339\%$ \\
8 & 0.82888 & $0.366\%$ \\
9 & 0.82881 & $0.373\%$ \\
10 & 0.82854 & $0.400\%$ \\
\bottomrule
\end{tabular}
\end{table}

\section{Related Work}

\paragraph{Feature interaction.}
Latent-factor collaborative filtering is the historical core of recommendation:
matrix factorization represents users and items as latent vectors whose
\emph{inner product} predicts affinity~\cite{koren2009}, and Factorization
Machines~\cite{rendle2010} generalized it to arbitrary multi-field
features---subsuming MF and its variants as special cases while retaining the
factorized generalization to unseen feature pairs; field-aware
variants~\cite{juan2016} refined the factorization. The deep era kept the inner
product central rather than replacing it: DeepFM~\cite{guo2017} runs an FM and
a DNN on shared embeddings, Wide\&Deep~\cite{cheng2016},
xDeepFM~\cite{lian2018}, DCN-v2~\cite{wang2021}, AutoInt~\cite{song2019}, and
FiBiNet~\cite{huang2019} extend explicit crosses in deeper or attention-based
forms. \textbf{Wukong}~\cite{zhang2024wukong} scales interaction
modeling with stacked factorization-machine and linear-compression blocks, and
\textbf{DHEN}~\cite{zhang2022dhen} composes heterogeneous interaction experts.

\paragraph{Sequential user-interest modeling.}
DIN~\cite{zhou2018} introduced target-aware attention pooling; DIEN, DSIN, and
BST~\cite{chen2019} added evolution and self-attention; long-sequence
methods such as SIM~\cite{pi2020}, ETA, TWIN~\cite{chang2023}, and
LONGER~\cite{chai2025longer}
retrieve or compress very long histories. Our \textbf{conditionally gated SwiGLU} is a
DIN-style mechanism: candidate context steers \emph{which positions matter}, but
enters only the gate, leaving the value path content-pure---a design that composes
cleanly with position-local compression.

\paragraph{Unifying feature interaction and sequence modeling.}
A fast-growing line of work brings the two branches into a single backbone.
\textbf{HSTU}~\cite{zhai2024} recasts
ranking/retrieval as a generative sequence task over a unified transformer.
\textbf{InterFormer}~\cite{zeng2024} interleaves sequence learning and
feature-interaction learning so the two refine each other layer by layer rather
than in sequence. \textbf{HyFormer}~\cite{hyformer} pairs a feature-interaction
module with a transformer that cross-attends behavioral history; our
sequence bus extends this to multiple domains plus a merged cross-domain stream,
routing explicit per-sequence dot products to the FM Highway (\S\ref{sec:perlayer}).
\textbf{OneTrans}~\cite{onetrans} and \textbf{TokenFormer}~\cite{tokenformer}
fuse all attributes, behaviors, and the target into one homogeneous
(decoder-only) stream; the latter must counter \emph{sequential collapse}, which
UniDot's two-bus separation avoids by construction. \textbf{UniMixer}~\cite{unimixer} and \textbf{TokenMixer
}~\cite{tokenmixer} pursue a single stackable token-mixing backbone (the latter
also probing scaling), \textbf{Kunlun}~\cite{kunlun} scales such ideas to
production, and Semantic-ID methods like \textbf{TIGER}~\cite{rajput2023} map
collaborative and content signals into a shared discrete-token space. The
TAAC $\times$ KDD Cup 2026 challenge~\cite{taac2026} formalizes this direction.
\textbf{SlimPer}~\cite{slimper2026} frames ranking as iterative refinement of a
compact, fixed-size $\langle$user, item$\rangle$ knowledge base that is re-read
against the full raw token set at every layer; UniDot is a preliminary
public-dataset test of this framework.

What distinguishes UniDot is its \emph{starting point}. Prior work unifies either
by scaling up feature mixing and treating behavior as more features~\cite{zeng2024,
unimixer, tokenmixer} or by embedding every feature as one more token in a
sequence model~\cite{zhai2024, onetrans, kunlun}; in both the
collaborative-filtering inner product becomes implicit. UniDot instead keeps it
explicit: cross-attention is read as FM scoring between query and key tokens, and
a dedicated \textbf{FM Highway} (\S\ref{sec:perlayer}) carries these low-order
interactions past the residual mixer~\cite{rendle2010}.

\section{Problem Formulation}\label{sec:problem}
We address \textbf{post-click conversion prediction}. An example is
$x=(u,i,\mathcal{S})$ with binary label $y\in\{0,1\}$ ($y{=}1$ denotes
conversion), where $u$ is the user profile (\texttt{user\_int} IDs, aligned
per-position weights, pre-trained user embeddings), $i$ the candidate item
(\texttt{item\_int} IDs, item embeddings), and $\mathcal{S}=\{S_s\}_{s=1}^{4}$ the
four behavioral domains. The model predicts a conversion probability and is trained
with binary cross-entropy. We now fix notation used throughout
\S\ref{sec:method}; we write $d$ for $d_{\text{model}}$.

\noindent\textbf{(1) Tokenization} maps raw inputs to $d$-dimensional tokens
(\S\ref{sec:tok}--\ref{sec:seqenc}):
\begin{equation}\label{eq:tok}
\begin{aligned}
&\mathbf{U}=\mathrm{Tok}_u(u)\in\mathbb{R}^{T_u\times d},\quad
\mathbf{I}=\mathrm{Tok}_i(i)\in\mathbb{R}^{T_i\times d},\\
&\mathbf{I}_h=\mathrm{Tok}_i^{h}(i)\in\mathbb{R}^{T_{ih}\times d},\quad
\mathbf{H}^{(s)}=\mathrm{Seq}(S_s)\in\mathbb{R}^{L_s\times d}.
\end{aligned}
\end{equation}

\noindent\textbf{(2) Stackable block.} Two token states are initialized from the
tokenizers, $Z^{0}_{\text{mix}}=[\mathbf{U};\mathbf{I}]$ (token-mixing bus) and
$Z^{0}_{\text{seq}}=\mathbf{I}_h$ (sequence-retrieval bus). For $\ell=1,\dots,L$
identical blocks:
\begin{equation}\label{eq:block}
\big(Z^{\ell}_{\text{mix}},\,Z^{\ell}_{\text{seq}},\,\phi^{\ell}\big)
=\mathrm{Block}_\ell\big(Z^{\ell-1}_{\text{mix}},\,Z^{\ell-1}_{\text{seq}},\,
\{\mathbf{H}^{(s)}\}_{s=1}^{S}\big),
\end{equation}
where $\phi^{\ell}$ is the layer's \textbf{FM Highway} signal (token dot-products
and Gram; \S\ref{sec:perlayer}).

\noindent\textbf{(3) Readout.} The classifier consumes a compressed readout
$\rho$ of the final states (\S\ref{sec:cls}), the highway signals of every
layer, and the skip-embedding signal $e_{\text{skip}}$ (\S\ref{sec:skip}):
\begin{equation}\label{eq:readout}
\hat{y}=\sigma\!\Big(\mathrm{MLP}\big[\,
\rho\big(Z^{L}_{\text{mix}},Z^{L}_{\text{seq}}\big);\,
\phi^{1};\dots;\phi^{L};\,
e_{\text{skip}}\,\big]\Big).
\end{equation}
The per-layer highway signals are \emph{concatenated}, not summed, so every
layer's explicit interactions reach the classifier undiluted.

\noindent\textbf{(4) Objective.} Binary cross-entropy plus an auxiliary delay
loss:
\begin{equation}\label{eq:obj}
\mathcal{L}=-y\log\hat{y}-(1-y)\log(1-\hat{y})+\lambda\,\mathcal{L}_{\text{delay}},
\end{equation}
where $\mathcal{L}_{\text{delay}}$ is an MSE regression of the log time-to-next-action
$\log\!\big(1+(t_{\text{label}}-t_{\text{event}})\big)$ over every row that has a
next action (clicks and conversions, not only positives), detailed in
\S\ref{sec:aux}.

\section{Method: UniDot}\label{sec:method}
\subsection{Overview}\label{sec:overview}
\paragraph{Unified tokenization.}
Every input---non-sequential user/item multi-field features \emph{and} per-position
behavioral-sequence events---is embedded into the same $d_{\text{model}}$ space and
represented as tokens (\S\ref{sec:tok}--\ref{sec:seqenc}). A single
\textbf{stackable macro-block} then processes them; stacking $L$ identical blocks
is the only depth knob, which makes the design simple to scale in depth and width.

The two buses co-evolve through $L$ macro-layers (each with $W$ token-mixing
sub-layers); the \textbf{MLP-Mixer fusion} exchanges information between them each
layer, and the \textbf{FM Highway} runs alongside (Figure~\ref{fig:arch}). A
shared sequence pipeline (\S\ref{sec:seqenc}) feeds both the cross-attention and a
per-layer \textbf{multi-channel pool} that injects pooled sequence tokens into the
non-sequential bus. We keep \S\ref{sec:method} symbolic
($L, W, T_u, T_i, T_{ih}, S, w, d$); concrete values are in \S\ref{sec:impl}
(Table~\ref{tab:config}).

\subsection{Tokenizers}\label{sec:tok}
Tokenization maps the model's \emph{heterogeneous} inputs (categorical fids,
multi-value ID lists, pre-trained embeddings, and behavioral sequences) into one
shared $d$-dimensional \textbf{token space} the macro-block can process
uniformly. Each multi-field input becomes a few $d$-dim tokens
($\mathrm{Tok}_u, \mathrm{Tok}_i$ in Eq.~\eqref{eq:tok}) by compression along the
token axis. Two primitives recur throughout the model, both mapping
$\mathbb{R}^{T\times d}\!\to\!\mathbb{R}^{T'\times d}$ with a LayerNorm tail:
\textbf{LCB}, a one-layer token-axis MLP-mixer (a single linear), and
\textbf{NCB}, its two-layer version (two linears with a GELU between).

\paragraph{Categorical (fid) features.}
Every fid is embedded \emph{per position} and the position bundle is compressed to
the token budget by a single learnable \textbf{NCB} along the token axis.
Because the NCB is \emph{learned}, the data decides which
feature combinations form each token, rather than a fixed pooling rule that
averages position-level salience away. This yields the user tokens $\mathbf{U}$
($T_u$) and, from two separate item tables, a compact view $\mathbf{I}$ ($T_i$,
token-mixing bus) and a richer $\mathbf{I}_h$ ($T_{ih}$, sequence-retrieval bus,
whose tokens drive the cross-attention queries). Behavioral sequences are
tokenized the same way (per-fid embedding then a \emph{position-local} fid-axis
NCB to a uniform per-position width; \S\ref{sec:seqenc}), once per forward and
shared by all consumers.

\paragraph{Multi-value fids: FAFE.}
A few high-value fids are \emph{lists} of behavioral IDs whose order carries no
meaning---they are \textbf{position-invariant}. For these, the static NCB is
replaced by a candidate-aware DIN-style attention pool against the ranking
candidate, so the field's token is a different combination of its values for each
candidate (\textbf{FAFE}, \S\ref{sec:fafe}): a special case of the same
compress-to-tokens step where the pooling weights are candidate-dependent.

\paragraph{Pre-trained embedding features.}
Dense pre-trained vectors (user SUM, LMF4Ads, \dots; item embeddings) are
normalized, projected by a small MLP into extra tokens, and appended to the
matching token set (\S\ref{sec:embproj}). Separately, a few paired
\texttt{user\_dense} arrays are not features but per-position multipliers on the
fid embeddings (\S\ref{sec:weights}).

\subsection{Sequence Encoder}\label{sec:seqenc}
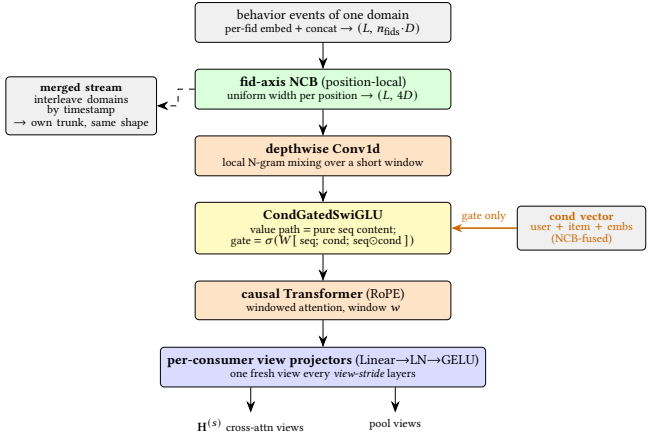
\begin{figure}[t]
\centering
\colorlet{condc}{orange!80!black}
\resizebox{\columnwidth}{!}{%
\begin{tikzpicture}[
  font=\footnotesize,
  >={Stealth[length=2mm]},
  stage/.style={draw, rounded corners=2pt, align=center, inner sep=3.5pt,
                minimum width=46mm},
  side/.style={draw, rounded corners=2pt, align=center, inner sep=3pt,
               fill=gray!12},
]
\node[stage, fill=gray!12] (inp) at (0,0)
  {behavior events of one domain\\[-1pt]
   {\scriptsize per-fid embed $+$ concat $\to (L,\, n_{\text{fids}}{\cdot}D)$}};
\node[stage, fill=green!14] (ncb) at (0,-1.2)
  {\textbf{fid-axis NCB} (position-local)\\[-1pt]
   {\scriptsize uniform width per position $\to (L,\, 4D)$}};
\node[stage, fill=orange!22] (dwc) at (0,-2.4)
  {\textbf{depthwise Conv1d}\\[-1pt]
   {\scriptsize local N-gram mixing over a short window}};
\node[stage, fill=yellow!30] (gate) at (0,-3.7)
  {\textbf{CondGatedSwiGLU}\\[-1pt]
   {\scriptsize value path $=$ pure seq content;}\\[-2pt]
   {\scriptsize gate $=\sigma(W[\,\text{seq};\,\text{cond};\,
    \text{seq}{\odot}\text{cond}\,])$}};
\node[side, text=condc, text width=21mm, align=center] (cond) at (4.65,-3.7)
  {\scriptsize \textbf{cond vector}\\[-2pt]\scriptsize user $+$ item $+$ embs\\[-2pt]
   \scriptsize (NCB-fused)};
\node[stage, fill=orange!22] (tfm) at (0,-5.0)
  {\textbf{causal Transformer} (RoPE)\\[-1pt]
   {\scriptsize windowed attention, window $w$}};
\node[stage, fill=blue!14] (views) at (0,-6.2)
  {\textbf{per-consumer view projectors} (Linear$\to$LN$\to$GELU)\\[-1pt]
   {\scriptsize one fresh view every \emph{view-stride} layers}};
\draw[->] (inp) -- (ncb);
\draw[->] (ncb) -- (dwc);
\draw[->] (dwc) -- (gate);
\draw[->] (gate) -- (tfm);
\draw[->] (tfm) -- (views);
\draw[->, condc, thick] (cond.west) -- node[above]{\scriptsize gate only} (gate.east);
\node[side, text width=25mm, align=center] (mrg) at (-4.35,-1.5)
  {\scriptsize \textbf{merged stream}\\[-1pt]
   \scriptsize interleave domains\\[-2pt]\scriptsize by timestamp\\[-1pt]
   \scriptsize $\to$ own trunk, same shape};
\draw[->, dashed] (ncb.west) -- ++(-0.3,0) |- (mrg.east);
\draw[->] (views.south) ++(-1.3,0) -- ++(0,-0.45)
  node[below]{\scriptsize $\mathbf{H}^{(s)}$ cross-attn views};
\draw[->] (views.south) ++(1.3,0) -- ++(0,-0.45)
  node[below]{\scriptsize pool views};
\end{tikzpicture}%
}
\caption{One sequence trunk (\S\ref{sec:seqenc}).}
\label{fig:trunk}
\end{figure}

The \textbf{sequence encoder} (the $\mathrm{Seq}(\cdot)$ of Eq.~\eqref{eq:tok})
takes the per-position sequence tokens from \S\ref{sec:tok} and turns them into the
views $\mathbf{H}^{(s)}$ consumed downstream, once per forward, shared by every
consumer. It is a pipeline of four stages, each chosen to model a different
structure in the behavior stream: a \textbf{cross-domain merge} (one timeline
across domains), a \textbf{depthwise conv} (local / N-gram structure), a
\textbf{DIN-style conditional SwiGLU} (candidate-aware filtering and enhancement),
and a \textbf{causal Transformer} (long-range token dependency).
Figure~\ref{fig:trunk} sketches one trunk.

\paragraph{Merged cross-domain sequence.}
A subset of domains is interleaved by timestamp into an extra cross-domain
stream, so the downstream module can learn cross-domain interactions more easily.
The real domains plus the merged stream give the downstream sequence count $S$.

\paragraph{Local mixing: depthwise Conv1d.}
A lightweight \textbf{depthwise Conv1d} mixes each channel over a local window,
cheaply capturing \textbf{N-gram patterns} (bursts, adjacent-event motifs) that
self-attention would otherwise spend capacity to relearn.

\paragraph{Information filtering: DIN-style conditional SwiGLU.}
A composite per-sample \textbf{cond vector} (user + item LCB tokens + emb-cond,
NCB-fused) enters \textbf{only the gate} of a SwiGLU; the value path stays a pure
function of the sequence. The gate selects positions by candidate relevance
(DIN-style) without altering their content.

\paragraph{Token dependency: causal Transformer.}
The global encoder is a shallow \textbf{Transformer} with RoPE and
\textbf{causal, windowed attention} (window $w$), modeling token dependencies in
time order through a fused-attention kernel (SDPA / FlashAttention). Per-consumer
\textbf{view projectors} ($\text{Linear}\to\text{LN}\to\text{GELU}$) then emit the
$D$-dim views $\mathbf{H}^{(s)}$, refreshed every view-stride layers.

\subsection{Per-layer computation}\label{sec:perlayer}
Each macro-layer runs the two buses in parallel and then fuses them.

\paragraph{Token-mixing bus.}
The bus state (the user, item, and pooled tokens concatenated) passes through
$W$ cross-token blocks. The block is a \textbf{swappable slot}; we use
\textbf{Wukong} (parallel \textbf{LCB + FMB} with residual, the FMB contributing
explicit pairwise dot-products). We also evaluated a TokenMixer-style
block~\cite{tokenmixer} and UniMixer~\cite{unimixer} in this slot, but neither
beat Wukong at our data scale---both likely need more data to converge. This is a
controlled slot-swap \emph{within} UniDot, not a matched end-to-end retrain of
those architectures (out of scope under the competition budget), so we read it as
indicative at our scale, not a definitive ranking. Pooled tokens are sliced off
between layers so the persistent state stays at $(T_u + T_i, D)$.

\begin{figure}[t]
\centering
\colorlet{chA}{blue!70}\colorlet{chB}{red!75}
\colorlet{chC}{green!55!black}\colorlet{chD}{orange!95!black}
\resizebox{\columnwidth}{!}{%
\begin{tikzpicture}[
  font=\footnotesize,
  >={Stealth[length=2mm]},
  itm/.style={draw, fill=gray!15, minimum size=4.2mm, inner sep=0pt},
  chan/.style={draw, thick, circle, inner sep=1.0pt, fill=white},
  ptok/.style={draw, thick, rounded corners=1.5pt, minimum width=7.5mm,
               minimum height=4mm, inner sep=1pt},
  box/.style={draw, rounded corners=2pt, align=center, inner sep=3.5pt},
]
\foreach \i/\x in {1/-3.0, 2/-2.0, 3/-1.0, 4/0.0, 5/1.0, 6/2.0}
  \node[itm] (h\i) at (\x,0) {};
\node at (2.65,0) {\scriptsize$\dots$};
\node[itm] (h7) at (3.3,0) {};
\node[anchor=east] at (-3.45,0) {\scriptsize items $h_\ell$};
\node[align=center] at (0.1,0.52) {\scriptsize
  $w_{\ell c}=\sigma\big(\mathrm{MLP}[\,h_\ell;\,
  \mathrm{summary}(Z_{\text{mix}})\,]\big)$\, (per-channel sigmoid gate)};
\node[chan, draw=chA] (sA) at (-2.25,-1.05) {$\Sigma$};
\node[chan, draw=chB] (sB) at (-0.75,-1.05) {$\Sigma$};
\node[chan, draw=chC] (sC) at (0.75,-1.05) {$\Sigma$};
\node[chan, draw=chD] (sD) at (2.25,-1.05) {$\Sigma$};
\foreach \p/\c/\w in {%
  1/A/0.90, 1/B/0.10,
  2/A/0.55, 2/B/0.35, 2/C/0.08,
  3/B/0.90, 3/A/0.10,
  4/B/0.30, 4/C/0.60, 4/D/0.08,
  5/C/0.85, 5/B/0.10,
  6/D/0.70, 6/A/0.20,
  7/D/0.90, 7/C/0.12}
  \draw[ch\c, opacity=\w, line width=0.4pt+\w pt] (h\p.south) -- (s\c.north);
\node[ptok, draw=chA, fill=chA!15] (pA) at (-2.25,-1.7) {\scriptsize $p_1$};
\node[ptok, draw=chB, fill=chB!15] (pB) at (-0.75,-1.7) {\scriptsize $p_2$};
\node[ptok, draw=chC, fill=chC!15] (pC) at (0.75,-1.7) {\scriptsize $p_3$};
\node[ptok, draw=chD, fill=chD!15] (pD) at (2.25,-1.7) {\scriptsize $p_4$};
\foreach \c in {A,B,C,D} \draw[->] (s\c) -- (p\c);
\node[box, fill=blue!10] (out) at (0,-2.45)
  {$\ell_2$-normalize each $\to$ concat over channels \& $S$ seqs $\to$ LayerNorm $\to$ bus};
\foreach \c in {A,B,C,D} \draw[->] (p\c.south) -- (out.north -| p\c);
\end{tikzpicture}%
}
\caption{MultiChannelSeqPool (\S\ref{sec:perlayer}): positions are gated into
$C$ channels by independent state-conditioned sigmoid weights.}
\label{fig:pool}
\end{figure}
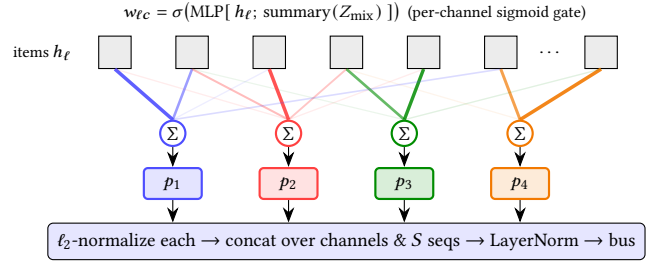

\paragraph{MultiChannelSeqPool.}
For each of the $S$ sequence views, a multi-channel pool---in the spirit of
NetVLAD / NeXtVLAD multi-cluster soft-assignment
aggregation~\cite{arandjelovic2016netvlad,lin2018nextvlad}---produces $C$ pooled
tokens. Each channel gates every position with an \textbf{independent sigmoid}
(not a softmax), conditioned on a summary of the current mixing-bus state, so
channels fire independently rather than competing for a fixed attention mass;
each pooled token is the gate-weighted sum of positions. Because sigmoid gating
leaves the magnitude unbounded, each pooled token is \textbf{$\ell_2$-normalized}
over $D$ to restore unit norm; the $S{\cdot}C$ tokens then pass a
\textbf{LayerNorm} tail before entering the bus, matching the NCB-tailed profile
tokens so the token-mix does not systematically down-weight them
(Figure~\ref{fig:pool}).

\paragraph{Sequence-retrieval bus.}
A single-history, single-layer instance of this bus recovers a standard
single-history target-attention CTR module~\cite{hyformer}; here we
generalize it to $S$ domains (real + merged), per-layer queries, and an
FM-Highway readout. With item state of $T_{ih}$ tokens:
\textbf{(i) per-sequence cross-attention}---the item state queries each of the
$S$ sequence views (no internal residual), giving raw outputs
$A_0,\dots,A_{S-1}\in\mathbb{R}^{B\times T_{ih}\times D}$;
\textbf{(ii) LCB aggregation}---$\text{attn\_agg}=\text{LCB}(\text{stack}(A_0,\dots))$,
linear because the fusion FFN below already supplies the nonlinearity;
\textbf{(iii) per-token fusion FFN}---a small FFN mixes each item token with that
token's retrieved vectors from all $S$ sequences into a $D$-dim residual delta on
the item state; and
\textbf{(iv) FM Highway signals} (bypass fusion, concatenated across layers)---each
layer contributes \emph{per-sequence dots} $d_i[t]=\langle\text{item}[t],A_i[t]\rangle$
(the per-token affinity between the candidate and each behavioral domain), a
learned \emph{fused-domain dot} (an NCB across the $S$ outputs, dotted with the
item state), the \emph{aggregated Gram} $G=\text{item}\cdot\text{attn\_agg}^\top$
(the full pairwise dot-product matrix), and, computed post-fusion,
\emph{user--item cross-dots} (inner products between the user tokens of the mixing
bus and the item tokens of the retrieval bus). Together these form $\phi^\ell$ in
Eq.~\eqref{eq:block}.

\paragraph{Bus-level fusion (\texttt{FuseFFN}, canonical MLP-Mixer~\cite{tolstikhin2021}).}
The three groups $w_{\text{out}}$, pooled, and $h_{\text{out}}$ are concatenated
along the token axis, then pass through an \textbf{NCB token-mix} (nonlinear
mixing across tokens), a token-wise \textbf{SwiGLU} channel-mix (applied to each
token, weights shared across tokens), and per-side
zero-initialized projections scaled by per-side learnable gates,
producing residual deltas on the two buses (Figure~\ref{fig:fuse}). Because
the side projections are zero-init, fusion starts as identity and is learned in.

\begin{figure}[t]
\centering
\colorlet{grpW}{orange!85!black}\colorlet{grpP}{green!55!black}
\colorlet{grpH}{blue!65}
\resizebox{0.75\columnwidth}{!}{%
\begin{tikzpicture}[
  font=\footnotesize,
  >={Stealth[length=2mm]},
  tok/.style={draw, minimum size=4.2mm, inner sep=0pt},
  mixb/.style={draw, rounded corners=2pt, align=center, inner sep=3.5pt,
               fill=yellow!30, minimum width=62mm},
  projb/.style={draw, rounded corners=2pt, align=center, inner sep=2.5pt,
                fill=gray!12},
  plus/.style={draw, circle, inner sep=0.6pt, thick},
]
\foreach \x in {-2.8,-2.3,-1.8,-1.3,-0.8}
  \node[tok, draw=grpW, fill=orange!22] at (\x,0) {};
\foreach \x in {-0.2,0.3,0.8}
  \node[tok, draw=grpP, fill=green!16] at (\x,0) {};
\foreach \x in {1.4,1.9,2.4,2.9}
  \node[tok, draw=grpH, fill=blue!13] at (\x,0) {};
\node[text=grpW] at (-1.8,0.52) {\scriptsize $w_{\text{out}}$ ($T_u{+}T_i$)};
\node[text=grpP] at (0.3,0.52)  {\scriptsize pooled ($S{\cdot}C$)};
\node[text=grpH] at (2.15,0.52) {\scriptsize $h_{\text{out}}$ ($T_{ih}$)};
\node[anchor=east] at (-3.25,0) {\scriptsize concat};
\node[mixb] (tmix) at (0.05,-0.75)
  {\textbf{NCB token-mix}: 2-layer MLP \emph{across the token axis} + LN\\
   {\scriptsize every token mixes with every other --- the cross-bus exchange}};
\node[mixb, fill=yellow!16] (cmix) at (0.05,-1.9)
  {\textbf{SwiGLU}: channel-mix over $D$, weights shared across tokens};
\draw[->] (0.05,-0.2) -- (tmix.north);
\draw[->] (tmix) -- (cmix);
\foreach \x in {-2.8,-2.3,-1.8,-1.3,-0.8}
  \node[tok, draw=grpW, fill=orange!22] at (\x,-2.7) {};
\foreach \x in {-0.2,0.3,0.8}
  \node[tok, draw=grpP, fill=green!16] at (\x,-2.7) {};
\foreach \x in {1.4,1.9,2.4,2.9}
  \node[tok, draw=grpH, fill=blue!13] at (\x,-2.7) {};
\draw[->] (cmix.south) -- (0.05,-2.5);
\node[text=gray, align=center] at (0.3,-3.3)
  {\scriptsize pool slice:\\[-2pt]\scriptsize no write-back};
\draw[gray, ->] (0.3,-2.95) -- (0.3,-3.1);
\node[projb, text=grpW] (pw) at (-1.8,-3.5)
  {\scriptsize $\mathrm{proj}_w$ (zero-init) $\times\, s_w$};
\node[projb, text=grpH] (ph) at (2.15,-3.5)
  {\scriptsize $\mathrm{proj}_h$ (zero-init) $\times\, s_h$};
\draw[->, grpW] (-1.8,-2.95) -- (pw.north);
\draw[->, grpH] (2.15,-2.95) -- (ph.north);
\node[plus, grpW] (aw) at (-1.8,-4.2) {$+$};
\node[plus, grpH] (ah) at (2.15,-4.2) {$+$};
\draw[->, grpW] (pw) -- node[right]{\scriptsize $\Delta_w$} (aw);
\draw[->, grpH] (ph) -- node[right]{\scriptsize $\Delta_h$} (ah);
\draw[->, grpW] (-3.1,0) -| (-3.6,-4.2) -- (aw.west);
\draw[->, grpH] (3.2,0) -| (3.75,-4.2) -- (ah.east);
\node[anchor=north] at (-1.8,-4.45) {\scriptsize $Z^{\ell}_{\text{mix}}$};
\node[anchor=north] at (2.15,-4.45) {\scriptsize $Z^{\ell}_{\text{seq}}$};
\end{tikzpicture}%
}
\caption{FuseFFN (\S\ref{sec:perlayer}).}
\label{fig:fuse}
\end{figure}

\subsection{Classifier readout}\label{sec:cls}
The readout is Eq.~\eqref{eq:readout}. The final bus states are \emph{not}
flattened wholesale. The compressed readout $\rho$ has two parts: (i) an NCB
compresses the $(T_u{+}T_i{+}T_{ih})$ concatenated bus tokens to a handful of
readout tokens,
which are flattened; and (ii) a \textbf{cross-dot gram}---all pairwise inner
products between the mixing-bus and retrieval-bus tokens, computed on the
\emph{uncompressed} states---is appended, so the compression cannot erase explicit
second-order signal. The classifier input concatenates $\rho$, the
layer-concatenated \textbf{FM Highway} $[\phi^{1};\dots;\phi^{L}]$ (per-sequence
dots, fused-domain dots, aggregated Grams, user--item cross-dots; LayerNorm'd),
and the LayerNorm'd \textbf{skip-embedding signal} $e_{\text{skip}}$
(\S\ref{sec:skip}). This feeds a 2-layer MLP and a linear head;
logits are clamped to $[-20, 20]$. An \textbf{auxiliary conversion-delay head}
(\S\ref{sec:aux}) adds a second loss term.

\subsection{Multi-path mutual learning}\label{sec:dml}
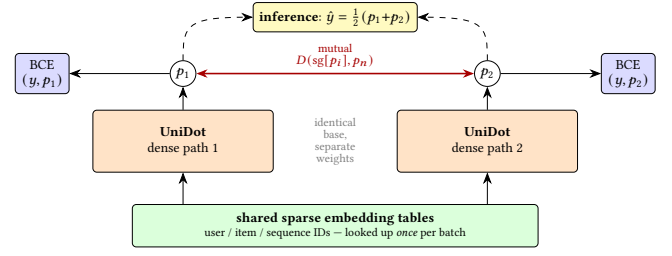
\begin{figure}[t]
\centering
\resizebox{\columnwidth}{!}{%
\begin{tikzpicture}[
  font=\footnotesize,
  >={Stealth[length=2mm]},
  emb/.style={draw, rounded corners=2pt, align=center, inner sep=4pt,
              fill=green!14, minimum width=72mm},
  pth/.style={draw, rounded corners=2pt, align=center, inner sep=4pt,
              fill=orange!22, minimum width=32mm, minimum height=11mm},
  prob/.style={draw, circle, inner sep=1.5pt, fill=white},
  loss/.style={draw, rounded corners=2pt, align=center, inner sep=3pt,
               fill=blue!14},
  serve/.style={draw, rounded corners=2pt, align=center, inner sep=3pt,
                fill=yellow!30},
]
\node[emb] (emb) at (0,0)
  {\textbf{shared sparse embedding tables}\\[-1pt]
   {\scriptsize user / item / sequence IDs --- looked up \emph{once} per batch}};
\node[pth] (p1) at (-2.7,1.5) {\textbf{UniDot}\\ dense path 1};
\node[pth] (p2) at ( 2.7,1.5) {\textbf{UniDot}\\ dense path 2};
\node[align=center, gray] at (0,1.5) {\scriptsize identical\\[-2pt]\scriptsize base,\\[-2pt]\scriptsize separate\\[-2pt]\scriptsize weights};
\draw[->] (emb.north -| p1.south) -- (p1.south);
\draw[->] (emb.north -| p2.south) -- (p2.south);
\node[prob] (q1) at (-2.7,2.7) {$p_1$};
\node[prob] (q2) at ( 2.7,2.7) {$p_2$};
\draw[->] (p1.north) -- (q1.south);
\draw[->] (p2.north) -- (q2.south);
\draw[<->, red!65!black, thick] (q1.east) --
  node[above, align=center]{\scriptsize mutual\\[-2pt]\scriptsize
  $D(\mathrm{sg}[p_i],p_n)$} (q2.west);
\node[loss] (l1) at (-5.2,2.7) {BCE\\$(y,p_1)$};
\node[loss] (l2) at ( 5.2,2.7) {BCE\\$(y,p_2)$};
\draw[->] (q1.west) -- (l1.east);
\draw[->] (q2.east) -- (l2.west);
\node[serve] (avg) at (0,3.7)
  {\textbf{inference}: $\hat y=\tfrac12(p_1{+}p_2)$};
\draw[->, dashed] (q1.north) .. controls +(0,0.6) and +(-1.2,0) .. (avg.west);
\draw[->, dashed] (q2.north) .. controls +(0,0.6) and +(1.2,0) .. (avg.east);
\end{tikzpicture}%
}
\caption{Multi-path mutual learning ($N{=}2$, \S\ref{sec:dml}).}
\label{fig:dml}
\end{figure}

We train a single model as \textbf{two} UniDot \emph{paths} jointly on the same
batches~\cite{zhang2018dml,yilmaz2024mutual}, sharing one set of (dominant) sparse
embeddings, and average their two logits at inference (Figure~\ref{fig:dml}). The
motivation is the regularization identified in the original \textbf{deep mutual
learning} work~\cite{zhang2018dml}: the two paths mimic each other's predictions
and converge to a \emph{wider, flatter minimum}---a robust solution they agree
on---and such flat minima generalize better to unseen data, exactly the property
our setting rewards, where the test period is one step ahead of training and most
candidate user--item pairs are new. We apply it here over a \emph{shared}
embedding table. Two paths already capture most of this gain at $2\times$ dense
cost; we kept that as our submission.

\paragraph{Mutual objective.}
Let $z_n$ be path $n$'s logit and $p_n=\sigma(z_n)$. Each path is supervised by
the task loss on its own prediction \emph{and} pulled toward the detached
predictions of the other paths:
\begin{equation}
\mathcal{L} = \sum_{n=1}^{N}\Big[\,\mathcal{L}_{\text{task}}(y, p_n)
  \;+\; \frac{\lambda}{N-1}\sum_{i\neq n} D\big(\,\mathrm{sg}[p_i],\, p_n\big)\Big],
\label{eq:dml}
\end{equation}
where $\mathcal{L}_{\text{task}}$ is the BCE (plus auxiliary delay head,
\S\ref{sec:aux}), $\mathrm{sg}[\cdot]$ is the stop-gradient, and
$D$ is a squared error on probabilities $D(a,b)=(a-b)^2$. The submission uses
$N{=}2$ paths, mutual weight $\lambda{=}20$, with the mutual term
switched on after the first epoch.

\paragraph{Serving.}
At inference we average the two paths' logits. Serving a \emph{single} path
instead costs $1\times$ and already scores $0.83184$ test AUC---only $0.033\%$
below the two-path mean ($0.83217$)---a viable cheap deployment, with the mean
recovering the small remainder.

\section{Experiments}
The dataset, setup, and implementation details are in
Appendices~\ref{sec:data}--\ref{sec:impl}; here we report results.

\subsection{Main results}
UniDot finished as the \textbf{runner-up} (Table~\ref{tab:leaderboard}), scoring
\textbf{0.83217} test AUC---driven by the unified dual-bus architecture with the
FM Highway (\S\ref{sec:method}) and by scaling it in depth, width, and multi-path
mutual learning (\S\ref{sec:scaling}). The final submission retrains on
\emph{all} data with EMA weights, peaking at epoch~5 (Table~\ref{tab:traincurve}).

\begin{table}[t]
\caption{Training dynamics by epoch of the $d{=}128$, $N{=}2$ multi-path run.
\emph{Eval}/\emph{Test (sel.)}: the model-selection run (\texttt{remote}, 10\%
held-out). \emph{Test (final)}: the \texttt{final} submission, retrained on
\emph{all} data with EMA. LL${=}$LogLoss (train over both paths, eval over their
mean); ``---'' = not submitted.}
\label{tab:traincurve}
\centering
\small
\setlength{\tabcolsep}{4pt}
\begin{tabular}{ccccccc}
\toprule
Ep. & Train AUC & Eval AUC & Test (sel.) & Test (final) & Train LL & Eval LL \\
\midrule
1 & 0.83193 & 0.84181 & 0.82772 & --- & 0.43395 & 0.21259 \\
2 & 0.84232 & 0.84441 & 0.83031 & 0.83036 & 0.42422 & 0.21113 \\
3 & 0.84606 & 0.84558 & 0.83136 & 0.83143 & 0.42072 & 0.21058 \\
4 & 0.84962 & 0.84607 & 0.83193 & 0.83195 & 0.41750 & \textbf{0.21029} \\
5 & 0.85485 & \textbf{0.84629} & \textbf{0.83196} & \textbf{0.83217} & 0.41288 & 0.21048 \\
6 & 0.86199 & 0.84568 & --- & 0.83212 & 0.40679 & 0.21116 \\
\bottomrule
\end{tabular}
\end{table}

\subsection{Incremental improvements}
Table~\ref{tab:ablation} traces test AUC from the competition baseline to our
submission, each row adding one change. The largest single jump is UniDot itself
($+1.10\%$); subsequent refinements (FM-Highway dots, depthwise conv, aux loss,
FAFE, EMA), dense scaling, and multi-path DML add the rest, $+1.82\%$ total.

\begin{table}[t]
\caption{Round-2 incremental improvements.}
\label{tab:ablation}
\centering
\small
\setlength{\tabcolsep}{4pt}
\begin{tabular}{llcc}
\toprule
\# & Change & Test AUC & $\Delta$ \\
\midrule
0 & competition baseline & 0.81398 & --- \\
1 & $+$ UniDot & 0.82500 & $+1.102$ \\
2 & $+$ item-id hash embedding & 0.82565 & $+0.064$ \\
3 & tune token counts & 0.82609 & $+0.044$ \\
4 & tune batch size & 0.82704 & $+0.095$ \\
5 & $+$ more FM-Highway dot products & 0.82722 & $+0.018$ \\
6 & $+$ depthwise-conv pre-trunk & 0.82736 & $+0.014$ \\
7 & $+$ auxiliary delay loss & 0.82812 & $+0.075$ \\
8 & $+$ FAFE & 0.82837 & $+0.026$ \\
9 & tune learning rate & 0.82894 & $+0.056$ \\
10 & $+$ EMA weights & 0.82993 & $+0.099$ \\
11 & scale $d_{\text{model}}$ 64$\to$96 & 0.83024 & $+0.031$ \\
12 & scale $d_{\text{model}}$ 96$\to$128 & 0.83043 & $+0.019$ \\
13 & $+$ multi-path DML ($d{=}64$) & 0.83128 & $+0.085$ \\
14 & multi-path DML ($d{=}128$) & 0.83196 & $+0.068$ \\
15 & $+$ all-data retrain & \textbf{0.83217} & $+0.021$ \\
\midrule
\multicolumn{2}{l}{total over baseline} & & $\mathbf{+1.818}$ \\
\bottomrule
\end{tabular}
\end{table}

\subsection{Component ablations}
Table~\ref{tab:component} removes individual components in \texttt{tiny} mode
(single-path, $d{=}64$, in-distribution held-out, so only \emph{relative} gaps
matter). Depth helps up to our default 6 macro-layers (8--10 overfit the 4M set),
and removing the \textbf{FM Highway} costs the most ($-0.127\%$); removing the
sequence cross-attention costs little ($-0.053\%$), as the multi-channel pool
captures much of the same signal.

Dropping the \textbf{FuseFFN} bus fusion slightly \emph{raises} AUC ($+0.022\%$)
but worsens LogLoss, so the module's contribution is not fully conclusive. We
read this not as evidence against cross-bus exchange but as an under-designed
fuser: its static routing is the weak link, and an input-conditioned
(second-order) fuser is the clearest open direction.

\begin{table}[t]
\caption{Component ablations (\texttt{tiny} mode, single-path, $d{=}64$;
in-distribution held-out eval). $\Delta$ is AUC vs.\ the full model.}
\label{tab:component}
\centering
\small
\setlength{\tabcolsep}{4pt}
\begin{tabular}{lccc}
\toprule
Variant & AUC $\uparrow$ & LogLoss $\downarrow$ & $\Delta$AUC \\
\midrule
Full UniDot (6 layers) & 0.83657 & \textbf{0.2151} & --- \\
\quad 2 macro-layers & 0.83600 & 0.2157 & $-0.057\%$ \\
\quad 4 macro-layers & 0.83603 & 0.2154 & $-0.054\%$ \\
\quad 8 macro-layers & 0.83624 & 0.2152 & $-0.033\%$ \\
\quad 10 macro-layers & 0.83615 & 0.2152 & $-0.042\%$ \\
\quad $-$ FM Highway (all) & 0.83530 & 0.2159 & $-0.127\%$ \\
\quad $-$ cross-bus dots only & 0.83570 & 0.2156 & $-0.087\%$ \\
\quad $-$ multi-channel seq pool & 0.83624 & 0.2155 & $-0.033\%$ \\
\quad $-$ token-mixing bus (identity) & 0.83590 & 0.2161 & $-0.067\%$ \\
\quad $-$ FuseFFN (no bus fusion) & \textbf{0.83679} & 0.2154 & $+0.022\%$ \\
\quad $-$ sequence cross-attention & 0.83604 & 0.2153 & $-0.053\%$ \\
\quad $-$ merged cross-domain stream & 0.83619 & 0.2152 & $-0.038\%$ \\
\bottomrule
\end{tabular}
\end{table}

\subsection{Scaling study}\label{sec:scaling}
On this task the productive scaling axis is \emph{dense} capacity, not sparse
(Table~\ref{tab:scaling}).
Doubling embedding width alone gave no A/B win---the high-cardinality tables are
not the bottleneck. Widening the dense path from $64$ to $96$ to $128$ improves
AUC monotonically with diminishing returns ($+0.050\%$ cumulative). A second
identical path under mutual distillation (\S\ref{sec:dml}) is a more effective
use of dense capacity: $+0.135\%$ at $d{=}64$, more than the entire width sweep,
doubling dense parameters and FLOPs but not the embedding tables. The two axes
compose: $d{=}128$ with two paths reaches $0.83196$ ($+0.203\%$ over the
$d{=}64$ single-path baseline), our submission; the all-data retrain then gives
$0.83217$.

\begin{table}[t]
\caption{Scaling study.}
\label{tab:scaling}
\centering
\small
\setlength{\tabcolsep}{4pt}
\begin{tabular}{lcccc}
\toprule
Configuration & Dense par. & GFLOP & Test AUC $\uparrow$ & $\Delta$ \\
\midrule
$d{=}64$, sparse fold$\times$2 & 18.8M & 5.9 & --- & no win \\
$d{=}64$ (baseline) & 18.8M & 5.9 & 0.82993 & --- \\
$d{=}96$ & 36.7M & 12.4 & 0.83024 & $+0.031\%$ \\
$d{=}128$ & 60.3M & 21.2 & 0.83043 & $+0.050\%$ \\
$d{=}64$, 2-path & 37.6M & 11.7 & 0.83128 & $+0.135\%$ \\
$d{=}128$, 2-path & \textbf{120.6M} & \textbf{42.5} & \textbf{0.83196} & $\mathbf{+0.203\%}$ \\
\bottomrule
\end{tabular}
\end{table}
\section{Conclusion}
We presented UniDot, a novel architecture that \textbf{unifies sequence
modeling and feature interaction} in a single stackable block: non-sequential
fields and multi-domain behavioral sequences become tokens in one shared space,
processed by a parallel token-mixing bus and a sequence-retrieval bus that
co-evolve and exchange deltas through a canonical MLP-Mixer each layer. Its
defining idea, the \textbf{FM Highway}, routes explicit
dot-product interactions around the fusion path straight to the classifier, keeping
factorization-machine-style second-order signal alive at negligible cost. UniDot
\textbf{finished as the runner-up on the Industrial track} of the TAAC $\times$ KDD Cup
2026 challenge, scoring \textbf{0.83217} AUC---driven by the unified architecture, by
scaling the block in depth and width, and by multi-path mutual learning over
shared embeddings, rather than by feature engineering. Future
work: deploying UniDot in a real production system and, at that scale, a
systematic scaling-law study of the unified block.

\bibliographystyle{ACM-Reference-Format}
\bibliography{refs}

@misc{taac2026,
  author       = {{TAAC \texttimes{} KDD Cup 2026}},
  title        = {Tencent Uni-Rec Challenge: Towards Unifying Sequence Modeling and Feature Interaction for Large-scale Recommendation},
  year         = {2026},
  howpublished = {Competition website},
  url          = {https://algo.qq.com}
}

@inproceedings{zhang2018dml,
  author    = {Zhang, Ying and Xiang, Tao and Hospedales, Timothy M. and Lu, Huchuan},
  title     = {Deep Mutual Learning},
  booktitle = {Proceedings of the IEEE Conference on Computer Vision and Pattern Recognition (CVPR)},
  year      = {2018},
  note      = {arXiv:1706.00384}
}

@misc{slimper2026,
  author        = {Wang, Siqi and Chen, Xianjie and Deng, Shaofeng and Chen, Albert and others},
  title         = {SlimPer: Make Personalization Model Slim and Smart},
  year          = {2026},
  eprint        = {2607.12281},
  archivePrefix = {arXiv},
  primaryClass  = {cs.IR}
}

@misc{yilmaz2024mutual,
  author        = {Y{\i}lmaz, {\.I}brahim Can and Aldemir, Said},
  title         = {Mutual Learning for Finetuning Click-Through Rate Prediction Models},
  year          = {2024},
  eprint        = {2406.12087},
  archivePrefix = {arXiv},
  primaryClass  = {cs.IR}
}

@inproceedings{zhang2024wukong,
  author    = {Zhang, Buyun and Luo, Liang and Liu, Xi and Li, Jay and Chen, Zeliang and Zhang, Weilin and Wei, Xiaohan and Hao, Yuchen and Tsang, Michael and Wang, Wenjun and Liu, Yang and Li, Huayu and Badr, Yasmine and Park, Jongsoo and Yang, Jiyan and Mudigere, Dheevatsa and Wen, Ellie},
  title     = {Wukong: Towards a Scaling Law for Large-Scale Recommendation},
  booktitle = {Proceedings of the 41st International Conference on Machine Learning (ICML)},
  year      = {2024}
}

@misc{zhang2022dhen,
  author = {Zhang, Buyun and Luo, Liang and Liu, Xi and others},
  title  = {{DHEN}: A Deep and Hierarchical Ensemble Network for Large-Scale CTR Prediction},
  year   = {2022},
  note   = {arXiv:2203.11014}
}

@inproceedings{zhou2018,
  author    = {Zhou, Guorui and Zhu, Xiaoqiang and Song, Chenru and Fan, Ying and Zhu, Han and Ma, Xiao and Yan, Yanghui and Jin, Junqi and Li, Han and Gai, Kun},
  title     = {Deep Interest Network for Click-Through Rate Prediction},
  booktitle = {Proceedings of the 24th ACM SIGKDD International Conference on Knowledge Discovery \& Data Mining (KDD)},
  year      = {2018}
}

@inproceedings{chen2019,
  author    = {Chen, Qiwei and Zhao, Huan and Li, Wei and Huang, Pipei and Ou, Wenwu},
  title     = {Behavior Sequence Transformer for E-commerce Recommendation in {Alibaba}},
  booktitle = {Proceedings of the 1st International Workshop on Deep Learning Practice for High-Dimensional Sparse Data (DLP-KDD)},
  year      = {2019}
}

@inproceedings{pi2020,
  author    = {Pi, Qi and Zhou, Guorui and Zhang, Yujing and Wang, Zhe and Ren, Lejian and Fan, Ying and Zhu, Xiaoqiang and Gai, Kun},
  title     = {Search-based User Interest Modeling with Lifelong Sequential Behavior Data for Click-Through Rate Prediction},
  booktitle = {Proceedings of the 29th ACM International Conference on Information and Knowledge Management (CIKM)},
  year      = {2020}
}

@inproceedings{chai2025longer,
  author    = {Chai, Zheng and Ren, Qin and Xu, Xijun and Chen, Hua and Zhang, Xiao and Hu, Zhongan and Liang, Tianxiang and Wu, Lei and Zhou, Lingzheng and Yu, Zhihao and Sun, Yang and Pan, Junwei},
  title     = {{LONGER}: Scaling Up Long Sequence Modeling in Industrial Recommenders},
  booktitle = {Proceedings of the 19th ACM Conference on Recommender Systems (RecSys)},
  year      = {2025},
  note      = {arXiv:2505.04421}
}

@inproceedings{chang2023,
  author    = {Chang, Jianxin and Zhang, Chenbin and Fu, Zhiyi and Zang, Xiaoxue and Guan, Lin and Lu, Jing and Hui, Yiqun and Leng, Dewei and Niu, Yanan and Song, Yang and Gai, Kun},
  title     = {{TWIN}: TWo-stage Interest Network for Lifelong User Behavior Modeling in CTR Prediction at {Kuaishou}},
  booktitle = {Proceedings of the 29th ACM SIGKDD Conference on Knowledge Discovery and Data Mining (KDD)},
  year      = {2023}
}

@inproceedings{zhou2019dien,
  author    = {Zhou, Guorui and Mou, Na and Fan, Ying and Pi, Qi and Bian, Weijie and Zhou, Chang and Zhu, Xiaoqiang and Gai, Kun},
  title     = {Deep Interest Evolution Network for Click-Through Rate Prediction},
  booktitle = {Proceedings of the 33rd AAAI Conference on Artificial Intelligence (AAAI)},
  year      = {2019},
  note      = {arXiv:1809.03672}
}

@inproceedings{feng2019dsin,
  author    = {Feng, Yufei and Lv, Fuyu and Shen, Weichen and Wang, Menghan and Sun, Fei and Zhu, Yu and Yang, Keping},
  title     = {Deep Session Interest Network for Click-Through Rate Prediction},
  booktitle = {Proceedings of the 28th International Joint Conference on Artificial Intelligence (IJCAI)},
  year      = {2019},
  note      = {arXiv:1905.06482}
}

@misc{chen2021eta,
  author        = {Chen, Qiwei and Pei, Changhua and Lv, Shanshan and Li, Chao and Ge, Junfeng and Ou, Wenwu},
  title         = {End-to-End User Behavior Retrieval in Click-Through Rate Prediction Model},
  year          = {2021},
  eprint        = {2108.04468},
  archivePrefix = {arXiv},
  primaryClass  = {cs.IR}
}

@inproceedings{zhou2024tin,
  author    = {Zhou, Haolin and Pan, Junwei and Zhou, Xinyi and Chen, Xihua and Jiang, Jie and Gao, Xiaofeng and Chen, Guihai},
  title     = {Temporal Interest Network for User Response Prediction},
  booktitle = {Companion Proceedings of the ACM Web Conference (WWW Companion)},
  year      = {2024},
  note      = {arXiv:2308.08487}
}

@inproceedings{zhai2024,
  author    = {Zhai, Jiaqi and Liao, Lucy and Liu, Xing and Wang, Yueming and Li, Rui and Cao, Xuan and Gao, Leon and Gong, Zhaojie and Gu, Fangda and He, Michael and Lu, Yinghai and Shi, Yu},
  title     = {Actions Speak Louder than Words: Trillion-Parameter Sequential Transducers for Generative Recommendations},
  booktitle = {Proceedings of the 41st International Conference on Machine Learning (ICML)},
  year      = {2024}
}

@inproceedings{zeng2024,
  author    = {Zeng, Zhichen and Liu, Xiaolong and Hang, Mengyue and Liu, Xiaoyi and others},
  title     = {{InterFormer}: Effective Heterogeneous Interaction Learning for Click-Through Rate Prediction},
  booktitle = {Proceedings of the 34th ACM International Conference on Information and Knowledge Management (CIKM)},
  year      = {2025}
}

@misc{hyformer,
  author = {Huang, Yunwen and Hong, Shiyong and Xiao, Xijun and Jin, Jinqiu and Luo, Xuanyuan and Wang, Zhe and Chai, Zheng and Wu, Shikang and Zheng, Yuchao and Lin, Jingjian},
  title  = {{HyFormer}: Revisiting the Roles of Sequence Modeling and Feature Interaction in CTR Prediction},
  year   = {2026},
  note   = {arXiv:2601.12681}
}

@misc{onetrans,
  author = {Zhang, Zhaoqi and Pei, Haolei and Guo, Jun and Wang, Tianyu and Feng, Yufei and Sun, Hui and Liu, Shaowei and Sun, Aixin},
  title  = {{OneTrans}: Unified Feature Interaction and Sequence Modeling with One Transformer in Industrial Recommender},
  year   = {2025},
  note   = {arXiv:2510.26104}
}

@misc{tokenformer,
  author = {Zhou, Yifeng and Hu, Yuehong and Feng, Zhixiang and Pan, Junwei and Wu, Kaihui and Li, Hanyong and Zhang, Shangyu and Huang, Shudong and Zhu, Zhangbin and Yin, Chengguo and Gu, Haijie and Jiang, Jie},
  title  = {{TokenFormer}: Unify the Multi-Field and Sequential Recommendation Worlds},
  year   = {2026},
  note   = {arXiv:2604.13737}
}

@misc{unimixer,
  author = {Ha, Mingming and Wang, Guanchen and Chen, Linxun and Rao, Xuan and Shi, Yuexin and Ma, Tianbao and Liu, Zhaojie and Fan, Yunqian and Lu, Zilong and Niu, Yanan and Li, Han and Gai, Kun},
  title  = {{UniMixer}: A Unified Architecture for Scaling Laws in Recommendation Systems},
  year   = {2026},
  note   = {arXiv:2604.00590}
}

@misc{tokenmixer,
  author = {Jiang, Yuchen and Zhu, Jie and Han, Xintian and Lu, Hui and Bai, Kunmin and Yang, Mingyu and Wu, Shikang and Zhang, Ruihao and Zhao, Wenlin and Bai, Shipeng and Zhou, Sijin and Yang, Huizhi and Liu, Tianyi and Liu, Wenda and Gong, Ziyan and Ding, Haoran and Chai, Zheng and Xie, Deping and Chen, Zhe and Zheng, Yuchao and Xu, Peng},
  title  = {{TokenMixer-Large}: Scaling Up Large Ranking Models in Industrial Recommenders},
  year   = {2026},
  note   = {arXiv:2602.06563}
}

@misc{kunlun,
  author = {Hou, Bojian and Liu, Xiaolong and Liu, Xiaoyi and Xu, Jiaqi and Badr, Yasmine and Hang, Mengyue and Chanpuriya, Sudhanshu and others},
  title  = {Kunlun: Establishing Scaling Laws for Massive-Scale Recommendation Systems through Unified Architecture Design},
  year   = {2026},
  note   = {arXiv:2602.10016}
}

@inproceedings{rajput2023,
  author    = {Rajput, Shashank and Mehta, Nikhil and Singh, Anima and Keshavan, Raghunandan and Vu, Trung and Heldt, Lukasz and Hong, Lichan and Tay, Yi and Tran, Vinh Q. and Samost, Jonah and Kula, Maciej and Chi, Ed H. and Sathiamoorthy, Maheswaran},
  title     = {Recommender Systems with Generative Retrieval},
  booktitle = {Advances in Neural Information Processing Systems (NeurIPS)},
  year      = {2023}
}

@inproceedings{wang2017dcn,
  author    = {Wang, Ruoxi and Fu, Bin and Fu, Gang and Wang, Mingliang},
  title     = {Deep \& Cross Network for Ad Click Predictions},
  booktitle = {Proceedings of the ADKDD'17},
  year      = {2017},
  note      = {arXiv:1708.05123}
}

@inproceedings{wang2021,
  author    = {Wang, Ruoxi and Shivanna, Rakesh and Cheng, Derek and Jain, Sagar and Lin, Dong and Hong, Lichan and Chi, Ed},
  title     = {{DCN V2}: Improved Deep \& Cross Network and Practical Lessons for Web-scale Learning to Rank Systems},
  booktitle = {Proceedings of the Web Conference (WWW)},
  year      = {2021}
}

@inproceedings{guo2017,
  author    = {Guo, Huifeng and Tang, Ruiming and Ye, Yunming and Li, Zhenguo and He, Xiuqiang},
  title     = {{DeepFM}: A Factorization-Machine based Neural Network for CTR Prediction},
  booktitle = {Proceedings of the 26th International Joint Conference on Artificial Intelligence (IJCAI)},
  year      = {2017}
}

@inproceedings{lian2018,
  author    = {Lian, Jianxun and Zhou, Xiaohuan and Zhang, Fuzheng and Chen, Zhongxia and Xie, Xing and Sun, Guangzhong},
  title     = {{xDeepFM}: Combining Explicit and Implicit Feature Interactions for Recommender Systems},
  booktitle = {Proceedings of the 24th ACM SIGKDD International Conference on Knowledge Discovery \& Data Mining (KDD)},
  year      = {2018}
}

@inproceedings{song2019,
  author    = {Song, Weiping and Shi, Chence and Xiao, Zhiping and Duan, Zhijian and Xu, Yewen and Zhang, Ming and Tang, Jian},
  title     = {{AutoInt}: Automatic Feature Interaction Learning via Self-Attentive Neural Networks},
  booktitle = {Proceedings of the 28th ACM International Conference on Information and Knowledge Management (CIKM)},
  year      = {2019}
}

@inproceedings{huang2019,
  author    = {Huang, Tongwen and Zhang, Zhiqi and Zhang, Junlin},
  title     = {{FiBiNET}: Combining Feature Importance and Bilinear Feature Interaction for Click-Through Rate Prediction},
  booktitle = {Proceedings of the 13th ACM Conference on Recommender Systems (RecSys)},
  year      = {2019}
}

@inproceedings{cheng2016,
  author    = {Cheng, Heng-Tze and Koc, Levent and Harmsen, Jeremiah and Shaked, Tal and Chandra, Tushar and Aradhye, Hrishi and Anderson, Glen and Corrado, Greg and Chai, Wei and Ispir, Mustafa and others},
  title     = {Wide \& Deep Learning for Recommender Systems},
  booktitle = {Proceedings of the 1st Workshop on Deep Learning for Recommender Systems (DLRS)},
  year      = {2016}
}

@inproceedings{rendle2010,
  author    = {Rendle, Steffen},
  title     = {Factorization Machines},
  booktitle = {Proceedings of the 10th IEEE International Conference on Data Mining (ICDM)},
  year      = {2010}
}

@inproceedings{juan2016,
  author    = {Juan, Yuchin and Zhuang, Yong and Chin, Wei-Sheng and Lin, Chih-Jen},
  title     = {Field-aware Factorization Machines for CTR Prediction},
  booktitle = {Proceedings of the 10th ACM Conference on Recommender Systems (RecSys)},
  year      = {2016}
}

@inproceedings{tolstikhin2021,
  author    = {Tolstikhin, Ilya and Houlsby, Neil and Kolesnikov, Alexander and Beyer, Lucas and Zhai, Xiaohua and Unterthiner, Thomas and Yung, Jessica and Steiner, Andreas and Keysers, Daniel and Uszkoreit, Jakob and Lucic, Mario and Dosovitskiy, Alexey},
  title     = {{MLP-Mixer}: An all-{MLP} Architecture for Vision},
  booktitle = {Advances in Neural Information Processing Systems (NeurIPS)},
  year      = {2021}
}

@misc{jordan2024,
  author       = {Jordan, Keller and Jin, Yuchen and Boza, Vlado and You, Jiacheng and Cesista, Franz and Newhouse, Laker and Bernstein, Jeremy},
  title        = {Muon: An optimizer for hidden layers in neural networks},
  year         = {2024},
  howpublished = {Blog post},
  url          = {https://kellerjordan.github.io/posts/muon/}
}

@misc{liu2025muon,
  author       = {Liu, Jingyuan and Su, Jianlin and Yao, Xingcheng and Jiang, Zhejun and Lai, Guokun and Du, Yulun and Qin, Yidao and Xu, Weixin and Lu, Enzhe and Yan, Junjie and others},
  title        = {Muon is Scalable for {LLM} Training},
  year         = {2025},
  eprint       = {2502.16982},
  archivePrefix = {arXiv},
  primaryClass = {cs.LG}
}

@inproceedings{lin2018nextvlad,
  author    = {Lin, Rongcheng and Xiao, Jing and Fan, Jianping},
  title     = {{NeXtVLAD}: An Efficient Neural Network to Aggregate Frame-level Features for Large-scale Video Classification},
  booktitle = {Proceedings of the European Conference on Computer Vision (ECCV) Workshops},
  year      = {2018}
}

@inproceedings{arandjelovic2016netvlad,
  author    = {Arandjelovi\'c, Relja and Gronat, Petr and Torii, Akihiko and Pajdla, Tom\'a\v{s} and Sivic, Josef},
  title     = {{NetVLAD}: CNN Architecture for Weakly Supervised Place Recognition},
  booktitle = {Proceedings of the IEEE Conference on Computer Vision and Pattern Recognition (CVPR)},
  year      = {2016}
}

@article{koren2009,
  author  = {Koren, Yehuda and Bell, Robert and Volinsky, Chris},
  title   = {Matrix Factorization Techniques for Recommender Systems},
  journal = {IEEE Computer},
  volume  = {42},
  number  = {8},
  pages   = {30--37},
  year    = {2009}
}
\appendix
\section{Dataset}\label{sec:data}
\subsection{Dataset}
The challenge releases a large-scale, fully anonymized advertising dataset built
from real Tencent ad logs, over two rounds; we compete on the
\textbf{Industrial track}. All sparse features are anonymized integer IDs and
all dense features are fixed-length float vectors; no raw content or PII is
released. Table~\ref{tab:data} summarizes the schema.

Each example is a user--item interaction with two kinds of inputs:
\begin{itemize}
\raggedright
\item \textbf{Non-sequential multi-field features}---\texttt{user\_int} /
\texttt{item\_int} categorical IDs (single- and multi-valued), and
\texttt{user\_dense} / \texttt{item\_dense} continuous vectors that include
\textbf{pre-trained embeddings} (e.g.\ SUM and LMF4Ads on the user side; item
embeddings on the item side).
\item \textbf{Behavioral sequence features} from \textbf{four domains} with
9--14 fields each (\texttt{seq\_a} to \texttt{seq\_d}): each domain is a
time-ordered list of events carrying timestamps and action types.
\end{itemize}

Two schema properties directly shape the model. First, several \texttt{user\_dense}
arrays are \textbf{element-aligned} with the matching \texttt{user\_int} arrays:
each dense value is a per-element statistic (e.g.\ a dwell time or score) for that
ID, not a standalone feature. UniDot consumes them as \textbf{per-position weights}
on the ID embeddings (\S\ref{sec:weights}), matching their intended semantics.
Second, a few IDs have extreme cardinality (fid 116 takes $\approx$9.4M
values), motivating the skip-embedding handling in \S\ref{sec:skip}.

\begin{table}[t]
\caption{Industrial-track dataset and schema (second round; first round in
parentheses).}
\label{tab:data}
\centering
\small
\begin{tabular}{ll}
\toprule
Quantity & Industrial track \\
\midrule
Train / test samples & 35M / 12M (2M, r1) \\
Columns (categories) & 142 (7) \\
User Int / Dense fids & 54 / 17 \\
Item Int / Dense fids & 17 / 4 \\
Pre-trained emb fids (u/i) & 7 / 4 \\
Aligned per-position fids & 10 \\
Behavioral domains & 4 (9/14/12/10 flds) \\
High-card.\ ID (fid 116) & $\approx$9.4M \\
Label / metric & Conversion / \textbf{AUC} \\
\bottomrule
\end{tabular}
\end{table}

\section{Experimental Setup}\label{sec:setup}
\textbf{Data:} TAAC $\times$ KDD Cup 2026 Industrial track (35M train / 12M
test; 2M in round 1), conversion label. \textbf{Metric:} \textbf{AUC} of ROC
(official); we also report LogLoss.
\textbf{Training:} effective batch $\approx$3k (local A/B) / $\approx$12k
(online, 4 GPUs), dual Adagrad/Muon (dense LR 4e-4, Muon weight decay 1e-3,
sparse LR 0.1), 100-step warmup, BCE + aux delay ($\lambda{=}0.01$), \texttt{torch.compile}
and bf16 autocast.
\textbf{Model:} submitted config in Table~\ref{tab:config} ($\approx$2.1B params,
$d_{\text{model}}{=}128$, $N{=}2$ multi-path).
\textbf{Inference:} the served model uses \texttt{torch.compile} and bf16
autocast. End-to-end inference plus evaluation on the 12M-example test
set---including the one-time compilation---takes $\approx$14{,}500\,s for the
two-path model and $\approx$7{,}200\,s for a single path (roughly half),
within the competition's inference budget. The bf16 autocast costs only a
slight AUC degradation ($<0.0001$).

\begin{table}[t]
\caption{Training/evaluation modes.}
\label{tab:modes}
\centering
\small
\setlength{\tabcolsep}{4pt}
\begin{tabular}{llll}
\toprule
Mode & Data (train / eval) & GPUs & Role \\
\midrule
\texttt{tiny}   & 4M / 1M, random hold-out      & 1       & rapid design A/B \\
\texttt{remote} & 35M / 10\% random hold-out    & 4 & full-scale validation \\
\texttt{final}  & 35M / no hold-out             & 4 & submission (EMA weights) \\
\bottomrule
\end{tabular}
\end{table}

\paragraph{Training and evaluation modes.}
We use three modes (Table~\ref{tab:modes}): \textbf{\texttt{tiny}} caps the data
to a random 4M-train / 1M-eval subset on one GPU for fast A/B of design choices;
\textbf{\texttt{remote}} promotes the surviving configuration to the full dataset
on 4 GPUs with a held-out split, confirming the win holds at full scale; and
\textbf{\texttt{final}} retrains on \emph{all} data and serves the EMA weights for
the submission.

\section{Implementation and Training Details}\label{sec:impl}
The components below are largely orthogonal to the core architecture
(\S\ref{sec:method})---data-specific feature handling and training machinery---so
we collect them here. They matter for reproducing the reported numbers.

\begin{table*}[t]
\caption{Submitted configuration behind the reported numbers.}
\label{tab:config}
\centering
\small
\begin{tabular}{ll}
\toprule
Group & Setting \\
\midrule
Macro-layers $L$ / mix-blocks-per-layer $W$ & 6 / 2 \\
$d_{\text{model}}$ & 128 \\
Token-mix block & Wukong (parallel LCB + FMB, rank 32) \\
User NS / user emb tokens & 8 / 2 ($\times$7 fids) $\Rightarrow T_u = 22$ \\
Item NS tokens (mix / retrieval bus) & 4 / 16 (+ 4 emb fids $\times$2) $\Rightarrow T_i = 12$, $T_{ih} = 24$ \\
Fuse / classifier hidden mult & 4 / 8 \\
Seq recent windows & a:256 b:256 c:512 d:512 \\
Merged stream & abcd:512 ($\Rightarrow S = 5$) \\
Seq fid-compress tokens & 4 ($\to$ 512-d/pos) \\
DIN cond tokens (u/i/emb/fused) & 2 / 4 / 1 / 4 \\
Trunk encoder & Transformer, 1 layer, 4 heads, causal, window $w{=}128$, RoPE \\
Pre-trunk encoder (per-dom.\ + merged) & depthwise Conv1d, kernel 21 \\
View projection & Linear$\to$LN$\to$GELU, view stride 1 \\
FuseFFN delta gates & sigmoid($\cdot$), init 0.5 \\
Classifier readout & NCB-compress to 4 tok + cross-dot gram \\
Skip-emb threshold / type / slots & 2M / hash / 2M \\
Item-id hash table & 2M slots \\
Dense optimizer & Muon (wd 1e-3); sparse: Adagrad \\
Loss & BCE; aux delay ($\lambda{=}0.01$) \\
Multi-path (DML) & $N{=}2$ paths, shared sparse emb \\
Mutual loss / weight & MSE / $\lambda{=}20$ (from epoch 1) \\
Params & $\approx$2.1B (embedding-dominated) \\
\bottomrule
\end{tabular}
\end{table*}

\subsection{Active configuration}\label{sec:activecfg}
Table~\ref{tab:config} resolves every symbol of \S\ref{sec:method} to its
submitted value. The token budgets work out to $T_u = 8 + 7{\cdot}2 = 22$
user-side tokens (8 NS tokens plus 2 per projected user-emb fid),
$T_i = 4 + 4{\cdot}2 = 12$ and $T_{ih} = 16 + 4{\cdot}2 = 24$ item-side tokens,
and $S = 4 + 1 = 5$ sequences (four behavioral domains plus the merged stream);
the fid-axis compression yields a uniform $4{\cdot}128 = 512$-d per-position
width. Parameters total $\approx$2.1B but are embedding-dominated (sparse
lookups); the dense compute path is tens of millions of parameters. Because the
per-fid embedding tables are $d_{\text{model}}$-wide, scaling $d_{\text{model}}$
from 64 to 128 roughly doubles the (dominant) embedding parameter count.

\subsection{Per-position weights (\texttt{user\_dense} pairing)}\label{sec:weights}
Ten \texttt{user\_dense} arrays of \S\ref{sec:data} are not features but
\textbf{per-position multipliers}. Each is aligned position-by-position with the
\emph{same-numbered} multi-value \texttt{user\_int} fid (matching length) and
scales that ID's embedding before pooling, so every value's contribution is
weighted by its paired statistic (e.g.\ dwell time or interaction count). Two
classes are handled differently. \textbf{Count-like} fids (\texttt{62}--\texttt{66},
\texttt{118}, \texttt{121}) pass through a \textbf{fixed} $\log(1+x)/10$
transform---the log tames the heavy tail, the $/10$ lands it in a unit-ish
range---and multiply the matching embedding. \textbf{Similarity-like} fids
(\texttt{89}--\texttt{91}, signed cosine / score values) are instead
\textbf{clamped to $[-1,1]$} and used directly as signed multipliers, so a
negative affinity can subtract a value's embedding. Both transforms use \emph{fixed} (non-learned) constants, so the
weighting is stable across the cold-restart re-initializations of
\S\ref{sec:opt}.

\subsection{Field-aware feature embedding (FAFE)}\label{sec:fafe}
\begin{figure}[t]
\centering
\resizebox{\columnwidth}{!}{%
\begin{tikzpicture}[
  font=\footnotesize,
  >={Stealth[length=1.8mm]},
  b/.style={draw, rounded corners=2pt, align=center, inner sep=3pt, minimum height=7mm},
  val/.style={b, fill=green!12},
  cand/.style={b, fill=blue!14},
  att/.style={b, fill=gray!12},
  sum/.style={b, fill=yellow!28},
  res/.style={b, fill=orange!22},
]
\node[cand] (q)    at (0,1.25)   {ranking candidate $\mathbf{e}_i$};
\node[val]  (vals) at (0,-0.15)  {field values\\ $\mathbf{e}_1,\dots,\mathbf{e}_K$};
\node[att]  (att)  at (4.2,0.55)  {DIN attention\\ $a_k=\operatorname{softmax}_k g(\mathbf{e}_i,\mathbf{e}_k)$};
\node[sum]  (sum)  at (8.3,0.55)  {weighted sum\\ $\mathbf{p}=\sum_k a_k\,\mathbf{e}_k$};
\node[res]  (out)  at (11.4,0.55) {candidate-aware\\ field token};
\draw[->] (q.east)    -- node[above,font=\scriptsize,pos=0.6]{query} (att.west);
\draw[->] (vals.east) -- node[below,font=\scriptsize,pos=0.6]{keys} (att.west);
\draw[->] (att) -- node[above,font=\scriptsize]{$a_k$} (sum);
\draw[->] (vals.south) -- ++(0,-0.5)
  -| node[below,font=\scriptsize,pos=0.25]{values $\mathbf{e}_k$} (sum.south);
\draw[->] (sum) -- (out);
\end{tikzpicture}%
}
\caption{Field-aware feature embedding (FAFE, \S\ref{sec:fafe}): a DIN-style
attention pool scores each value of a multi-value field against the ranking
candidate, yielding a candidate-dependent field token; all other fields keep the
static NCB pooling.}
\label{fig:fafe}
\end{figure}
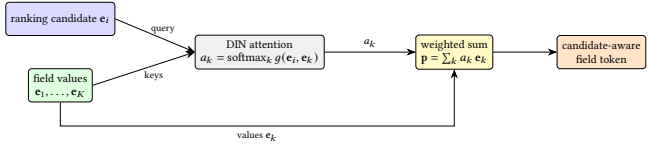

A few high-value \emph{multi-value} \texttt{user\_int} fields---fids \texttt{15},
\texttt{63}--\texttt{66}, \texttt{115}--\texttt{118}, \texttt{121}, \texttt{122}
(lists of behavioral IDs)---are pooled \textbf{candidate-aware} instead of
statically: a DIN-style attention pool scores each value in the list against the
ranking candidate
(Fig.~\ref{fig:fafe}), so the field's pooled embedding is a \emph{different
combination of its values for each candidate}. The remaining fields use the
static NCB pooling of \S\ref{sec:tok}. This concentrates candidate-awareness on
the large behavioral ID lists where it pays off, without the regression of making
the whole tokenizer target-aware.

\subsection{Pre-trained embedding projection and normalization}\label{sec:embproj}
Pre-trained dense vectors (user: SUM, LMF4Ads, \dots; item embeddings) enter on
their own path. \textbf{Normalization.} Their heavy-tailed \emph{count}
dimensions are first standardized to zero mean / unit variance with \textbf{fixed
statistics} estimated once on the training set (not per batch), so inference sees
the same scaling and the projector is not destabilized by the long tail.
\textbf{Projection.} Each normalized vector is mapped by a 2-layer
(Linear--GELU--Linear) MLP into a few $d$-dim tokens, which a \textbf{shared
per-token LayerNorm} then puts at unit scale---matching the
LayerNorm-tailed NCB tokens (\S\ref{sec:tok}) so the bus token-mix does not
systematically down-weight the embedding tokens. The resulting tokens are appended
to the user / item token sets.

\paragraph{Structured sub-vector embeddings.}
One pre-trained fid (LMF4Ads, $320$-d) is not a single vector but a concatenation
of $10$ independently \textbf{L2-normalized} $32$-d sub-vectors, each a unit
vector or all-zeros (padding). Flattening it into one $320$-d input and applying
the MLP above would blur this block structure and mix the separately-normalized
sub-spaces, so this fid gets a \textbf{structured tokenizer}: reshape to
$10\times 32$, lift each $32$-d slot to a token with a \emph{shared} linear map,
mask out the zero-padded slots (present iff the slot norm is nonzero, so an empty
slot's bias cannot leak in), then compress the $10$ slot tokens with an LCB to the
same per-fid token count as the flat path (its LayerNorm tail giving unit-variance
tokens). All other emb fids (e.g.\ SUM, a single whole-vector-normalized $256$-d
embedding) keep the flat MLP projection.

\subsection{High-cardinality skip embeddings}\label{sec:skip}
Fids above a threshold (2M) are taken off the per-fid path and handled by a shared
\texttt{hash} (hashing-trick) table. Their pooled signal also feeds the classifier directly
($e_{\text{skip}}$ in Eq.~\eqref{eq:readout}), and they are re-initialized on cold
restart. The item-id fid additionally gets a dedicated 2M-slot
multiplicative-hash table whose embedding feeds the item tokenizers.

\subsection{Auxiliary conversion-delay head}\label{sec:aux}
A dedicated trunk + head regresses $\log(1+(t_{\text{label}}-t_{\text{event}}))$,
the delay to the user's next action, under an MSE loss. The mask is
$t_{\text{label}}>t_{\text{event}}$ (a next action exists), so it covers
\emph{both} next-click-without-conversion and conversion rows---roughly all rows
rather than the $\sim$12\% positives, $\sim$8$\times$ more aux signal. It
contributes the $\lambda\,\mathcal{L}_{\text{delay}}$ term of Eq.~\eqref{eq:obj}
($\lambda{=}0.01$).

\subsection{Optimization and training}\label{sec:opt}
\textbf{Dual optimizer.} Embedding parameters use \textbf{Adagrad} (sparse
gradients). Dense parameters are split: the $\geq$2D \emph{matrix} weights use
\textbf{Muon}~\cite{jordan2024} (\textbf{Moonshot variant}~\cite{liu2025muon},
which adds decoupled weight decay and rescales the orthogonalized update to match
AdamW's update RMS so AdamW-tuned learning rates transfer with no separate Muon LR
search; weight decay 1e-3), while the 1D parameters (LayerNorm scales, biases) go
to an \textbf{AdamW auxiliary} group (weight decay 0). Both share a linear LR
warmup, with optional decay at each cold-restart boundary.
\textbf{Cold restart:} at every epoch boundary after the first, embedding tables
above a cardinality threshold---\emph{all} tables in the active config
(threshold 0)---are re-initialized together with their Adagrad state, while dense
parameters persist: the backbone trains across epochs while embeddings are
re-learned each epoch, a regularizer matched to the one-step-ahead train/test
time gap. After each re-initialization the dense learning rate is \textbf{warmed
up afresh} (a second warmup schedule), letting the re-learned embeddings settle
before full-rate updates resume. \textbf{Time bucketing:} continuous deltas
$t_{\text{now}} - t_{\text{event}}$ map to 64 embedding slots (0 = padding),
boundaries spanning $\approx$1\,s to $\approx$1.5\,years, with capacity
concentrated in the 1\,h--18\,month range (recency-weighted grid).
\textbf{Logit clamp:} forward logits are clamped to $[-20,20]$.
\textbf{EMA weights:} we track an exponential moving average of the dense
parameters (decay $0.999$); when a held-out split is available we publish, each
epoch, whichever of the live or EMA weights scores higher on it, and the final
all-data submission serves the EMA weights. The EMA is \textbf{reset} to the live
weights over each post-reinit warmup window, so the high-variance re-warm steps
do not contaminate the average.

\section{Data-scaling study}\label{sec:datascaling}
\begin{figure}[t]
\centering
\begin{tikzpicture}
\begin{axis}[
  width=0.95\columnwidth, height=5.1cm,
  xmode=log, log basis x=2,
  xtick={4,8,16,32}, xticklabels={4M,8M,16M,32M},
  xmin=3.4, xmax=42,
  xlabel={training examples (\texttt{tiny} mode)},
  axis y line*=left,
  ylabel={held-out AUC}, ylabel style={blue!65!black},
  ymin=0.8350, ymax=0.8460,
  ytick={0.8360,0.8380,0.8400,0.8420,0.8440},
  scaled y ticks=false,
  yticklabel style={/pgf/number format/fixed, /pgf/number format/fixed zerofill, /pgf/number format/precision=4, blue!65!black},
  every tick/.append style={blue!65!black},
  grid=both, grid style={gray!16},
  tick label style={font=\scriptsize},
  label style={font=\footnotesize},
]
\addplot[blue!65!black, mark=*, line width=1pt] coordinates {(4,0.83657)(8,0.83899)(16,0.84187)(32,0.84396)};
\end{axis}
\begin{axis}[
  width=0.95\columnwidth, height=5.1cm,
  xmode=log, log basis x=2, xmin=3.4, xmax=42,
  axis x line=none,
  axis y line*=right,
  ylabel={LogLoss}, ylabel style={red!70!black},
  ymin=0.2108, ymax=0.2158,
  ytick={0.2120,0.2135,0.2150},
  scaled y ticks=false,
  yticklabel style={/pgf/number format/fixed, /pgf/number format/fixed zerofill, /pgf/number format/precision=4, red!70!black},
  every tick/.append style={red!70!black},
  tick label style={font=\scriptsize},
  label style={font=\footnotesize},
]
\addplot[red!70!black, mark=square*, line width=1pt] coordinates {(4,0.2151)(8,0.2139)(16,0.21247)(32,0.21145)};
\end{axis}
\end{tikzpicture}
\caption{Data scaling (\texttt{tiny} mode, single-path $d{=}64$): held-out AUC
(left, \textcolor{blue!65!black}{blue}) and LogLoss (right,
\textcolor{red!70!black}{red}) vs.\ training examples ($\log_2$ axis). Both
improve log-linearly across the full 4M--32M range.}
\label{fig:datascaling}
\end{figure}
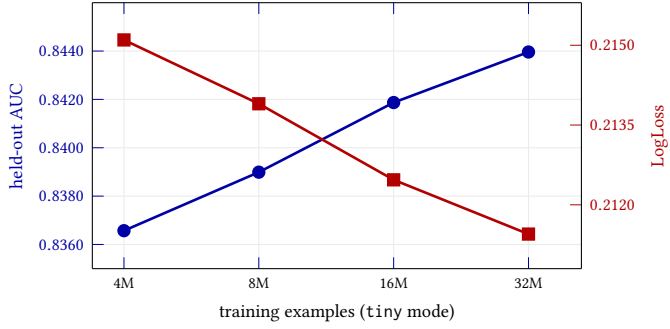

Complementing the model-scaling study (\S\ref{sec:scaling}), we grow the
\emph{training data} while holding the model at the \texttt{tiny} configuration
(single-path $d{=}64$) and changing only the training-example cap. Held-out AUC
rises \textbf{log-linearly} with data---$0.83657$ (4M), $0.83899$ (8M),
$0.84187$ (16M), $0.84396$ (32M), about $+0.0025$ AUC per doubling---and LogLoss
falls in step ($0.2151$, $0.2139$, $0.2125$, $0.2115$); see
Figure~\ref{fig:datascaling}. The clean, unbroken log-linear trend through the
full 32M suggests the unified block is data-hungry and has not saturated,
motivating the production / scaling-law follow-up.

\end{document}